\documentclass[amssymb,amsmath,pre,twocolumn,aps,showpacs]{revtex4}
\usepackage{amsmath}
\usepackage{graphicx}
\usepackage{dcolumn}

\usepackage[dvips]{epsfig}

\begin{document}
\title{Using self-driven microswimmers for particle separation} 
\author{W.~Yang$^{1,2}$}
\author{V.~R.~Misko$^{1}$}
\author{K.~Nelissen$^{1,3}$}
\author{M.~Kong$^{4}$}
\author{F.~M.~Peeters$^{1,3}$}

\affiliation{
\it $^{1}$Departement Fysica, Universiteit Antwerpen, Groenenborgerlaan 171, B-2020 Antwerpen, Belgium \\
\it $^{2}$College of Materials Science and Engineering, Taiyuan University of Science and Technology, Taiyuan 030024, P. R. China \\
\it $^{3}$Departamento de F\'isica, Universidade Federal do Cear\'a, Caixa Postal 6030, Campus do Pici, 60455-760 Fortaleza, Cear\'a, Brazil \\ 
\it $^{4}$Institute of Plasma Physics, Chinese Academy of Sciences, Hefei 230031, P. R. China 
}

\date{\today}

\begin{abstract} 
Microscopic self-propelled swimmers capable of autonomous navigation through complex environments provide appealing opportunities for localization, pick-up and delivery of micro-and nanoscopic objects. 
Inspired by motile cells and bacteria, man-made microswimmers have been fabricated, and their motion in patterned surroundings has been experimentally studied. 
We propose to use self-driven artificial microswimmers 
for separation of binary mixtures of colloids. 
We revealed different regimes of separation including one with a velocity inversion. 
Our finding could be of use for various biological and medical applications. 
\end{abstract}


\maketitle

\section{Introduction}

Quasi-one-dimensional~(Q1D) colloidal systems play an increasingly important role as model systems to study a variety of collective phenomena in low dimensional condensed matter physics.
The well-controlled way in real space and time, and the tunable interparticle interaction potential of colloidal systems lead to insights of a wide range of systems.
Most of the studies on Q1D colloidal systems focused on stripe structures and crystal 
defects~\cite{DoylePRE70,RicciPRE74,DoyleLgM22,StrattonPRE79}, 
solid-liquid phase transitions~\cite{DoylePRE72,FerreiraJPCM22}, 
single-file diffusion~\cite{LeidererScie287,BechingerPRL93,TaloniPRL96,KwintenEPL80} 
and transport behavior under external driving 
force~\cite{PiacentePRB72,LeidererPRL97,MiskoPRE80,ReichhardtPRE83}, 
including, e.g., dragged particles~\cite{dragged}. 
Transport properties of colloidal particles driven by an external force in narrow channels became a growing topic during the last decade.  
This is by part because the transport behavior might provide a deeper understanding of several biological systems, 
e.g., ion-channel transport 
in cell membranes~\cite{AstumianPRL91,RothPRL95}, 
DNA manipulations and 
separations~\cite{HanScie288,DoyleScie295,DoyleMacro44}. 
Also it is a readily available physical model to mimic other types of particles, 
including different ions and electrons, 
which might lead to practical use in microdevices and nanodevices~\cite{SiwyPRL89,HeJACS131}.  
Moreover, the transport behaviors are important in real systems of medicine, 
food production and field-response materials as well.      
There are also studies of transport properties of particles which are driven in other shapes of patterned 
environments~\cite{GrierPRL92,NoriPRB74}. 

In nature and technology, many systems are mixtures of different particle types. 
The distinction and interaction, mixture and separation of each particle type give rise to an exceedingly rich phenomenology as  compared to monodisperse systems. 
This motivited a number of studies on complex mixtures 
and, in particular, 
binary mixtures (see, e.g., 
\cite{ssnm2002,ssfmfn,ssfmfnPRE2004,ssfmfnPRE2005,ssfmfn2004,wyangPRE79}). 

One of the fundamental problems related to binary mixtures is their separation (see, e.g.,~
\cite{ssnm2002,ssfmfn,ssfmfnPRE2004,ssfmfnPRE2005,ssfmfn2004}). 
This problem is commonly tackled by using different asymmetric potentials for separating, e.g., interacting binary mixtures driven on periodic substrates~
\cite{ssnm2002,ssfmfn,ssfmfnPRE2004,ssfmfnPRE2005,ssfmfn2004}). 
or ferrofluids~\cite{engel}. 
Mixtures of particles can be driven either along the symmetry axis of the potentials, or in the transverse 
direction~\cite{ssPRB2004} 
which was demonstrated in the context of separation of macromolecules, DNA, or even cells 
(see, e.g.,~\cite{ertas,huangnb,berger,boxer,huangsc}). 

Note that in the previous studies a driving force 
(e.g., gravity~\cite{LeidererPRL97} or 
external field~\cite{PiacentePRB72,MiskoPRE80,ReichhardtPRE83}) was applied to {\it all} 
the particles of the system 
(or to one of the species), and the separation occurred either due to the different dynamic 
response of the 
species to the driving/potentials or due to their 
different self-diffusion constant. 

In contrast to that, here we propose a new mechanism of particle separation. 
It is based on injecting 
special particles which are able to move in the binary mixture and interact with its species. 
These can be either particles driven by external forces (which describes a typical situation in microrheology~\cite{Gnann}) 
or {\it self-driven} particles 
i.e., driven by 
various self-phoretic forces 
produced by a chemical (self-chemophoresis) 
(see, e.g., the recent review paper~\cite{Ebbens}), 
electrical (self-electrophoresis) or thermal 
(self-thermophoresis) gradient 
that the particle generates around itself 
(see Ref.~\cite{Buttinoni} and references therein). 
For example, so-called Janus particles 
($\mu$m-sized particles covered by gold on one 
of the hemispheres) can move (``swim'') 
when being illuminated by light 
(details of the mechanism of motion of these artificial 
``microswimmers'' (MS) are beyond the scope of this work and 
can be found in~\cite{Buttinoni,ArXivSwimmer}). 
Due to their amphliphilicity, Janus particles are extensively studied as a model of structure-directing amphiphiles 
(see, e.g.,~\cite{chen}) 
and in relation to their applications in new functional materials 
(see, e.g.,~\cite{SynytskaAFM,SynytskaTex}).  
Here we demonstrate that artificial MS, 
when moving in a binary mixture, 
are capable of {\it selectively driving} 
and separating the flows of different species of the binary mixture.

\section{Simulation} 

We consider a binary system of paramagnetic colloidal particles 
confined in a 2D infinitely long narrow hard-wall channel 
in an external perpendicular magnetic field. 
The particles 
(including the microswimmer --- for experimental realization of paramagnetic microswimmers see Refs.~\cite{Kline,Tierno}) 
interact via a repulsive dipole-dipole potential 
(see, e.g., experiments~\cite{DoyleLgM22,LeidererPRL97})
$V_{ij}(\vec{r}_{i},\vec{r}_{j})=Q_{i}Q_{j}/|\vec{r}_{i}-\vec{r}_{j}|^{3}$, 
where $Q_{i}=M_{i}\sqrt{\mu_{0}/4\pi}$ is the ``effective charge'', $\vec{M}_{i}$ is the magnetic moment, and 
$\vec{r}_{i}$ is the coordinate of the $i$-th particle; 
$\mu_{0}$ is the magnetic permittivity. 
We assume that 
there are two types of particles, A and B, which differ by size and charge. 
We call them ``small'' (A) and ``big'' (B) particles. 
The system consists of $N_{A}$ small particles with $Q_{A}$, 
and $N_{B}$ big particles with $Q_{B}=8 Q_{A}$. 
The effective charge of self-driven MS is chosen $Q_{d}=Q_{A}$. 
We introduce a unit energy $E_{0}=Q_{A}^{2}/a_{0}^{3}$ 
of the inter-particle interaction 
$V_{ij}(\vec{r}_{i},\vec{r}_{j})$, 
where the unit length $a_{0}=10$~$\mu m$ is on the order of the average distance 
between particles~\cite{DoyleLgM22,LeidererPRL97}. 
Thus the inter-particle interaction can be characterized by the dimensionless 
coupling parameter $\Gamma=Q_{A}^{2}/a_{0}^{3}k_{B}T$, 
where $k_{B}$ is the Boltzmann constant and $T$ the ambient temperature. 
In our simulations, we choose $\Gamma=200$ which corresponds to 
the strongly correlated regime, i.e., below the melting transition. 
Correspondingly, the unit force is $F_{0}=\Gamma k_{B}T/a_{0}$. 
For the chosen $\Gamma=200$ and room temperature, $F_{0}=8.28\cdot 10^{-14}$~N. 

The motion of colloidal particles is investigated using the Langevin equations of motion in the overdamped regime, i.e., the Brownian dynamics~(BD) method. 
This approach neglects hydrodynamic interactions as well as the 
short-time momentum relaxation of the particles 
(see, e.g.,~\cite{LeidererPRL97,MiskoPRE80,KwintenEPL74}). 
The overdamped equations of motion of particles are: 
\begin{equation}
\frac{d\vec{r}_{i}}{dt}=\frac{D_{i}}{k_{B}T}\{-\nabla_{\vec{r}_{i}}\sum_{i\neq j}V_{ij}(\vec{r}_{i},\vec{r}_{j})+\vec{F}_{drx}+\tilde{F}_{i}(t)\},
\label{BD}
\end{equation}
where $D_{i}$ is the self-diffusion coefficient, which is chosen as 
$D_{i}^{A}=1.0$~$\mu m^2/s$ for type A (and for MS) and 
$D_{i}^{B}=0.7$~$\mu m^2/s$ for type B particles, 
according to the experiment~\cite{LeidererPRL97,ArXivSwimmer}. 
The first term of the rhs of Eq.~(\ref{BD}) is the sum of all repulsive interaction forces acting on each particle, 
the second term is the ``self-driving force'' 
acting {\it only} on the MS 
(or ``motor'' particle~\cite{BradyPRL,BradyPRL2009}). 
Recently, a similar approach has been used for the description 
of motion of swimming bacteria and self-driven 
particles~\cite{bacteria}. 
Note that force $\vec{F}_{drx}$ entering into the Langevin equation (\ref{BD}) 
is an effective force which describes the propulsion mechanism 
on average (see, e.g., the recent studies~\cite{Lowen1,Lowen2} 
on the Brownian motion of a self-propelled particle). 
In experiment, this effective ``driving force'' can be tuned by, e.g., changing the intensity of the incident light (in case of Janus particles), the geometry of the experiment~\cite{ArXivSwimmer}, or the number of MS particles, etc.; we do not discuss here the relation between $\vec{F}_{drx}$ and those parameters. 
Furthermore, without loss of generality we consider here only the $x$-component of $\vec{F}_{dr}$. 
This is justified by the fact that the (average) {\it direction} of motion of a MS (i.e., a Janus particle illuminated by light) can be controlled by, e.g., direction of the incident light and by using different illumination patterns~\cite{Buttinoni,ArXivSwimmer}, 
or by additionally applied drift force~\cite{ArXivSwimmer}. 
Furthermore, in our case the motion of a MS across the channel is restricted by the boundaries. 
For simplicity, we ignore rotational motion of a 
MS~\cite{rotation}. 
The last term is the thermal random force that obeys the following conditions: 
$ 
\langle\tilde{F}_{i}(t)\rangle=0, 
$ 
and 
$ 
\langle\tilde{F}_{i\alpha}(t)\tilde{F}_{j\beta}(t')\rangle = 
2 \left( k_{B}^2 T^2 / D_{i} \right) \delta(t-t')\delta_{ij}\delta_{\alpha\beta}.
$

As a simulation cell, we use a unit box of length 
$L_{x} = 100$ ($1 mm$) and width $L_{y} = 12$ ($120 \mu m$) 
(the box dimensions are close to those in the 
experiments~\cite{DoyleLgM22,LeidererPRL97}). 
Periodic boundary conditions~(PBC) are imposed in the $x$ direction, and hard wall boundaries in the $y$ direction. 
We use a cutoff~($r = 6$) for the dipole short-range inter-particle interaction. 
We also checked that by doubling the length of the simulation cell our results do not change. 
     
\begin{figure}
\begin{center}
\includegraphics[width=0.96\columnwidth]{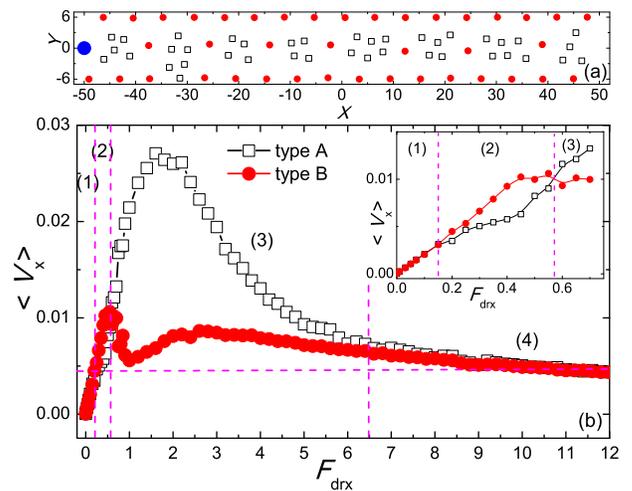}\\
\caption{
(a) Equilibrium configuration of a system with 
equal number of particles of the two species 
$N_{A}=40$, $N_{B}=40$ and 
$Q_{d}=8Q_{A}$. 
(b) The average velocity of particles 
of type A, $\langle V_{x}^{A} \rangle$ 
(black open squares connected by solid lines) 
and B, $\langle V_{x}^{B} \rangle$ 
(red dots connected by lines), 
as a function of $F_{drx}$. 
The microswimmer (MS) is shown by a blue filled circle. 
The same symbols are used in Figs. 2 to 5. 
The inset shows a zoom of the inversion velocity region (2). 
}
\label{vwithf} 
\end{center}
\end{figure}

The initial equilibrium configurations of the system were calculated for $\vec{F}_{drx}=0$. 
Typical equilibrium distribution is shown in Fig.~\ref{vwithf}(a) 
for $N_{A} = N_{B} = 40$ and $N_{MS} = 1$, 
which is consistent with our previous results~\cite{wyangPRE79} 
(without MS particle; note that in our present simulations 
$N_{MS} \ll N_{A}, \ N_{B}$). 
Then we calculated the average velocities of A and B particles as a function of $\vec{F}_{drx}$. 
Simulations were typically performed for $10^8$ time steps 
$\Delta t=10^{-4}$~(s).

\section{Separation of the particle flows} 

The average velocity (along the channel) $\langle V_{x} \rangle$ 
of particles A and B versus 
$\vec{F}_{drx}$
is shown in Fig.~\ref{vwithf}(b) for a typical system with 
$N_{A} = N_{B}=40$. 
The main characteristic feature of the function 
$\langle V_{x} \rangle(F_{drx})$ 
is a well-pronounced particle flow separation for a broad range 
of $F_{drx}$. 
Note that particles A (small) move {\it faster} than their counterpart. 
Depending on $F_{drx}$, 
we can distinguish four different regions 
(shown in Fig.~\ref{vwithf}(b) as (1) to (4)). 

\smallskip

\noindent
{\it Region (1): Rigid body motion} 
 
\noindent
This regime corresponds to very small $F_{drx}$ when the system 
is in a ``rigid body'' motion, i.e., velocities 
$\langle V_{x}^{A} \rangle$ and 
$\langle V_{x}^{B} \rangle$ coincide. 
The system is in a solid state ($\Gamma=200$), and 
only elastic deformations are present~\cite{MiskoPRE80}. 

\smallskip

\noindent
{\it Region (2): Inverse velocity motion} 

\begin{figure}
\begin{center}
\includegraphics[width=0.96\columnwidth]{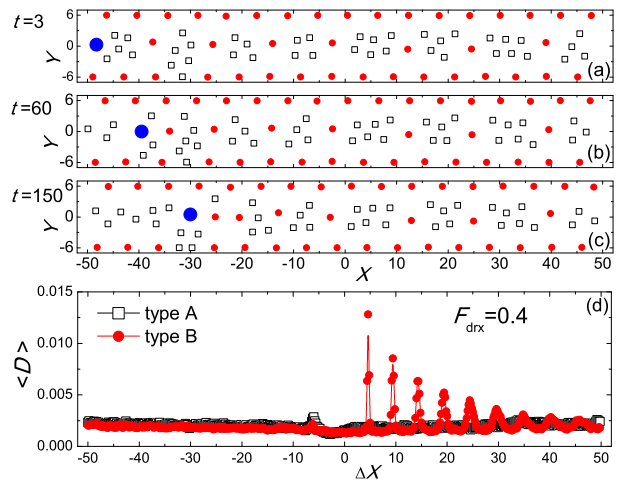}\\
\caption{
(a-c) Snapshots of the configurations of the binary system 
(small particles of type A are shown by empty black squares; 
large particles of type B are shown by red dots) 
in presence of a microswimmer (shown by a blue filled circle), 
for 
$N_{A}=40,\ N_{B}=40, \ Q_{d}=8Q_{A}$ 
and $F_{drx}=0.4$, which corresponds to Regime (2) 
(see Fig.~1) of inverse velocities. 
The snapshots are recorded at 
$t=3$ (a), 60 (b), and 150 (c). 
(d) The local density distribution of the particles versus 
distance to the MS, $\Delta X$. 
}
\label{TrajXdens0.4}
\end{center}
\end{figure}

\noindent
This regime is observed for small $F_{drx}$ when 
the velocity of the MS is high enough to locally melt the 
solid, however, the velocites are inversed as compared to the 
broad region (3), i.e., 
particles A (small) move {\it slower} than particles B. 
This unusual behavior is explained by the stronger interaction 
of the MS with particles B than with particles A. 
As a consequence, particles B are carried along by the MS 
while the dynamical friction due to the particle motion 
is very small in this case and can be neglected. 
This regime is further illustrated in 
Fig.~\ref{TrajXdens0.4}. 
The snapshots 
(Figs.~\ref{TrajXdens0.4}(a-c)) 
show that the system adjusts itself to the slow ``swimming'' 
of the MS, without appreciable changes in the structure. 
However, 
the local density profile 
(Fig.~\ref{TrajXdens0.4}(d)) 
indicates that 
the local density of particles B in front of the moving MS 
is higher than that of particles A. 
This means that particles B (big) are carried along by the MS 
while particles A (which interact weaker with the MS) ``slip'' 
and thus move slower. 

\smallskip

\noindent
{\it Region (3): Strong flow separation} 

\begin{figure}
\begin{center}
\includegraphics[width=0.96\columnwidth]{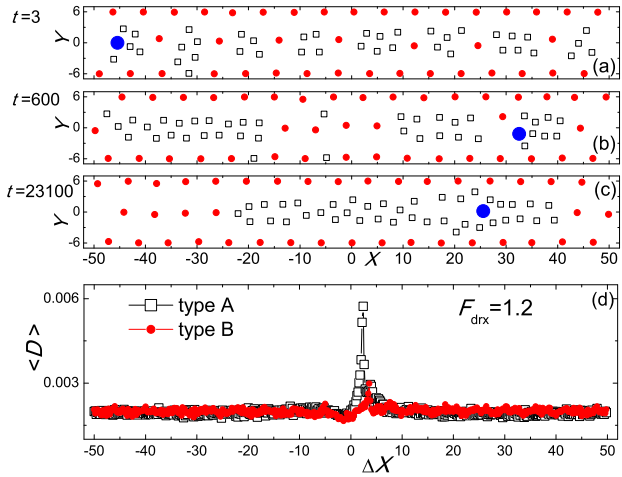}\
\caption{
(a-c) Snapshots of the configurations of the binary system 
in presence of a microswimmer 
(the same symbols are used as in Fig.~2), 
for 
$N_{A}=40,\ N_{B}=40, \ Q_{d}=8Q_{A}$ 
and $F_{drx}=1.2$, which corresponds to Regime (3) 
(see Fig.~1) of strong flow separation. 
The snapshots are recorded at 
$t=3$ (a), 600 (b), and 23100 (c). 
(d) The local density distribution of the particles versus 
distance to the MS, $\Delta X$. 
}
\label{TrajXdens1.2}
\end{center}
\end{figure}

\noindent
This regime is observed for a broad range of intermediate $F_{drx}$: 
this is the main and the most robust regime. 
With increasing velocity, the dynamical friction becomes 
increasingly important, and, as a result, type A particles 
(which are characterized by a larger self-diffusion coefficient) 
start to move faster. 
The function $\langle V_{x}^{A} \rangle(F_{drx})$ 
reaches its maximum at about 
$F_{drx} = 2$ 
and then gradually decreases 
while 
the function $\langle V_{x}^{B} \rangle(F_{drx})$ 
increases only slowly in this region and then decreases. 
It is worth noting that in this regime, 
the function $\langle V_{x}^{B} \rangle(F_{drx})$ first 
rapidly decreases. 
This decrease is accompanied by a simultaneous sharp 
increase in 
the function $\langle V_{x}^{A} \rangle(F_{drx})$. 
The origin of this behavior is explained by the fact that 
the motion of the two species (i.e., particles A and B) 
becomes less correlated as compared to that in regime (2). 
This loss of correlation in motion of the different species 
resembles the transition from elastic to plastic motion. 
The observed splitting of the average velocities is similar 
to the splitting in the velocities of adjacent layers found 
in a ``vortex Wigner solid''~\cite{VRMCorbino,VRMCorbino1} 
where the transition to plastic motion was induced by shear 
stress. 
The snapshots and the local density profiles shown in  
Fig.~\ref{TrajXdens1.2} 
illustrate this regime. 
In contrast to regime (2) 
(cp. Fig.~\ref{TrajXdens0.4}), 
the motion of the MS strongly influences the particle 
distribution: in particular, we observe a pronounced 
effect of particle separation in real space 
(i.e., not only in velocity space), 
or the formation of {\it clusters} of particles A 
which ``trap'' the MS and follow it 
(see Figs.~\ref{TrajXdens1.2}(b, c)). 
The local density of particles A has a sharp peak in front 
of the MS (Fig.~\ref{TrajXdens1.2}(d)), 
due to the MS-induced compression of the particle configuration 
in front of the MS. 
At the same time, the particles behind the MS are relaxed 
resulting in no peak in the local density. 

\smallskip

\noindent
{\it Region (4): Fast MS motion.} 

\noindent
This regime corresponds to large $F_{drx}$: 
finally, the system undergoes a transition to a 
``quasi-rigid body'' regime when the MS moves too fast 
through the binary solid to produce a responce in form 
of flow separation. 

\smallskip 

In order to analyze the role of the width of the channel, 
we performed simulations for wider channels, and we found 
that the effect of separation remains qualitatively similar: 
for a broad range of $F_{dr}$, the average velocity of small 
particles, 
$\langle V_{x}^{A} \rangle(F_{drx})$, 
is larger than that of large particles, 
$\langle V_{x}^{B} \rangle(F_{drx})$. 
However, the more gentle effect of velocity inversion can 
disappear, depending on the concentration of the particles. 
The role of imbalance in properties of the particles 
is discussed in the next section.

\section{Role of imbalance in particle properties} 

Above, we revealed the general features of the separation of the 
particle flows 
by the MS ``swimming'' in a binary mixture and explained the mechanisms 
of the observed effect for the different separation regimes. 
Here we analyze the influence of the imbalance in particle properties, 
i.e., differences in charge of the MS and the constituent particles 
and the relative density of particles, 
on the observed effect.

\begin{figure}
\begin{center}
\includegraphics[width=0.96\columnwidth]{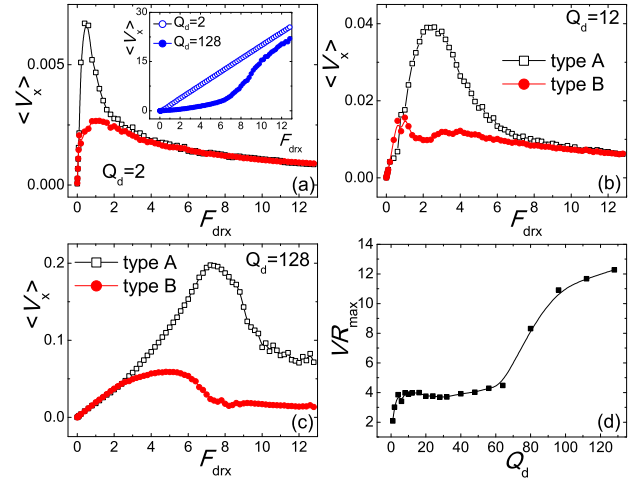}\\
\caption{
(a-c) 
The average velocity $\langle V_{x} \rangle$ of particles A and B 
as function of $F_{drx}$ for a system  with $N_A=N_B=40$ and 
different charge of the MS, 
$Q_d=2Q_A$ (a), $12Q_A$ (b), and $128Q_A$ (c). 
The inset in (a) shows the velocity of the driven aprticle versus $F_{drx}$ for systems with $Q_d=2Q_A$ and $Q_d=128Q_A$. 
(d) The maximum ratio of the velocities of particles A and B, 
$VR_{max}$, as a function of charge of the MS, $Q_d$. 
}
\label{VRwithQd}
\end{center}
\end{figure}

We systematically examined systems with the charge of the MS 
varying in a broad range from 
$Q_d=Q_A$ to $Q_d=128Q_A$. 
The results of our calculation of 
$\langle V_{x}\rangle$ versus $F_{drx}$ 
for 
$Q_{d}=2Q_A$, 12$Q_A$, and 128$Q_A$ are presented in 
Figs.~\ref{VRwithQd}(a-c). 
These results show that the position of the maximum average velocity 
$\langle V_{x}^{A}\rangle$ shifts towards larger $F_{drx}$ 
with increasing the $Q_d/Q_A$ ratio. 
We explain this by the fact that increasing the interparticle interaction 
enables the system response to the motion of the MS 
even for large velocities, 
with simultaneous extension of the initial rigid-body region 
(see Fig.~\ref{VRwithQd}(c)). 
Note that the region of inverted velocities exists only for 
a certain range of values of $Q_d/Q_A \sim 10$. 
 
We analyzed the maximum effect of flow separation, 
i.e., the ratio 
$\langle V_{x}^{A}\rangle / \langle V_{x}^{B}\rangle \equiv VR_{max}$ 
and calculated its dependence on $Q_d/Q_A$ (Fig.~\ref{VRwithQd}(d)): 
the initial fast grow is followed by a wide ``plateau'' 
(for $Q_d/Q_A = 10$ to 60) where 
$\langle V_{x}^{A}\rangle / \langle V_{x}^{B}\rangle 
\approx 4$, 
then followed by a considerable increase for 
$Q_d/Q_A > 60$. 

\begin{figure}
\begin{center}
\includegraphics[width=0.96\columnwidth]{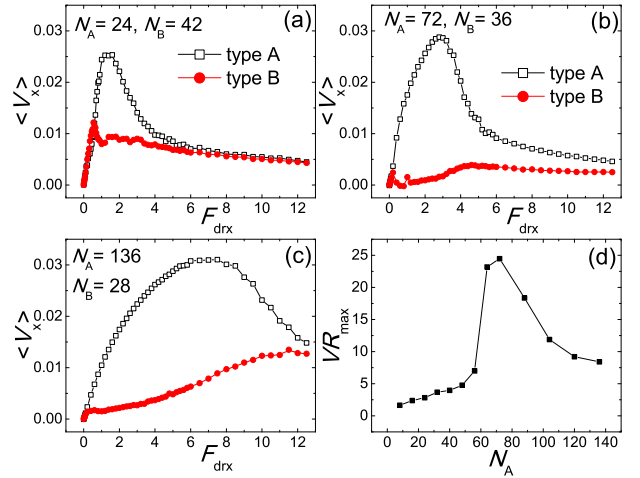}\\
\caption{
(a-c) The function $\langle V_{x} \rangle$ for particles A and B 
versus the driving force for a system with $Q_d=8Q_A$ and different 
relative densities of particles A and B. 
(d) The maximum ratio of the velocities of particles A and B, 
$VR_{max}$, as a function of number of particles A, $N_A$.
}
\label{VRwithNs}
\end{center}
\end{figure}

We also studied the role of the imbalance in the density of 
particles A and B on the flow separation. 
We found that 
the position of the maximum of 
$\langle V_{x}^{A}\rangle$ shifts towards larger $F_{drx}$ 
with increasing the $N_{A}/N_{B}$ ratio. 
At the same time, 
the maximum effect of flow separation, 
$\langle V_{x}^{A}\rangle / \langle V_{x}^{B}\rangle$ 
turns out to be a non-monotonic function of 
$N_{A}/N_{B}$, 
with a maximum at $N_{A}/N_{B} \approx 2$ 
(see Fig.~\ref{VRwithNs}).

\section{Conclusions}

We demonstrated that particles in a binary mixture can be effectively separated by using self-driven particles, or microswimmers, that move through the system. 
We analyzed the main features of the particle separation and explained mechanisms of different regimes 
including one with a velocity inversion. 
Our theoretical findings can be verified in experiments with microswimmers in colloidal binary mixtures. 
Indeed, the revealed regimes of particle flow separation are based on a rather general interplay between the interaction and the dynamical friction. 
In addition, in our model we used realistic experimental parameters. 
For example, our estimates for the ``driving force'' needed to 
observe the predicted behavior is on the order of (0.1-0.2)pN 
which agrees with the estimates for light-driven Janus particles 
in the experiments~\cite{Buttinoni,ArXivSwimmer}. 
Furthermore, the same experimental studies reported on: 
(i) the high control of the direction of motion of microswimmers 
(i.e., by employing various illumination patterns) and 
(ii) the tunability of the ``driving force'' of light-driven Janus particles (i.e., by tuning the intensity of the illumination). 
These two ingredients are of key importance for the possibility of experimental realization of the present theoretical findings. 
Moreover, the approach used in this work is also applicable to microrheology. 
The results of our study can be useful for mixture separation in biology and medicine.

\section{Acknowledgments}

This work was supported by the ``Odysseus'' Program of the Flemish Government and the Flemish Science Foundation (FWO-Vl) (Belgium), 
by the National Natural Science Foundation of China (No. 11047111), 
the State Key Program of National Natural Science of China (No. 51135007), the Research Fund for the Doctoral Program of Higher    Education of China (No. 20111415120002), and the Major State Basic Research Development Program of China (973) (No. 2009CB724201).


\begin{thebibliography}{99}

\bibitem{DoylePRE70} 
R. Haghgooie and P. S. Doyle, Phys. Rev. E~{\bf 70}, 061408~(2004).

\bibitem{RicciPRE74}  
A. Ricci, P. Nielaba, S. Sengupta, and K. Binder, 
Phys. Rev. E~{\bf 74}, 010404(R)~(2006).

\bibitem{DoyleLgM22}  R. Haghgooie, C. Li, and P. S. Doyle, Langmuir~{\bf 22}, 3601~(2006).

\bibitem{StrattonPRE79} 
T. R. Stratton, S. Novikov, R. Qato, S. Villarreal, B. Cui, 
S. A. Rice, and B. Lin, 
Phys. Rev. E~{\bf 79}, 031406~(2009).

\bibitem{DoylePRE72} 
R. Haghgooie and P. S. Doyle, Phys. Rev. E~{\bf 72}, 011405~(2005).

\bibitem{FerreiraJPCM22} 
W. P. Ferreira, G. A. Farias and F. M. Peeters, J. Phys.: Condens. Matter~{\bf 22}, 285103~(2010).

\bibitem{LeidererScie287} 
Q.-H. Wei, C. Bechinger, and P. Leiderer, Science~{\bf 287}, 625~(2000).

\bibitem{BechingerPRL93} 
C. Lutz, M. Kollmann, and C. Bechinger, Phys. Rev. Lett.~{\bf 93}, 026001~(2004).

\bibitem{TaloniPRL96} 
A. Taloni and F. Marchesoni, Phys. Rev. Lett.~{\bf 96}, 020601~(2006).

\bibitem{KwintenEPL80} 
K. Nelissen, V. R. Misko and F. M. Peeters, EuroPhys. Lett.~{\bf 80}, 56004~(2007).

\bibitem{PiacentePRB72} 
G. Piacente and F. M. Peeters, Phys. Rev. B~{\bf 72}, 205208~(2005).

\bibitem{LeidererPRL97} 
M. K\"{o}ppl, P. Henseler, A. Erbe, P. Nielaba, and 
P. Leiderer, 
Phys. Rev. Lett.~{\bf 97}, 208302~(2006).

\bibitem{MiskoPRE80} 
D. V. Tkachenko, V. R. Misko, and F. M. Peeters, Phys. Rev. E~{\bf 80}, 051401~(2009).

\bibitem{ReichhardtPRE83} 
C. Reichhardt, C. Bairnsfather, and C. J. Olson Reichhardt, Phys. Rev. E~{\bf 83}, 061404~(2011).

\bibitem{dragged} 
C. Bairnsfather, C. J. Olson Reichhardt, and C. Reichhardt, EuroPhys. Lett.~{\bf 94}, 18001~(2011). 

\bibitem{AstumianPRL91} 
R. D. Astumian, Phys. Rev. Lett.~{\bf 91}, 118102~(2003).

\bibitem{RothPRL95} 
R. Roth and D. Gillespie, Phys. Rev. Lett.~{\bf 95}, 247801~(2005).

\bibitem{HanScie288} 
J. Han and H. G. Craighead, Science~{\bf 288}, 1026~(2000).

\bibitem{DoyleScie295} 
P. S. Doyle, J. Bibette, A. Bancaud, and J.-L. Viovy, Science~{\bf 295}, 2237~(2002).

\bibitem{DoyleMacro44} 
D. W. Trahan and P. S. Doyle, 
Macromolecules~{\bf 44}, 383~(2011).

\bibitem{SiwyPRL89} 
Z. Siwy and A. Fuli\'{n}ski, Phys. Rev. Lett.~{\bf 89}, 198103~(2002).

\bibitem{HeJACS131} 
Y. He, D. Gillespie, D. Boda, I. Vlassiouk, R. S. Eisenberg, 
and Z. S. Siwy, 
J. Am. Chem. Soc.~{\bf 131}, 5194~(2009).

\bibitem{GrierPRL92} 
A. Gopinathan and D. G. Grier, Phys. Rev. Lett.~{\bf 92}, 130602~(2004).

\bibitem{NoriPRB74} 
S. Savel'ev, A. L. Rakhmanov, and F. Nori, Phys. Rev. B~{\bf 74}, 024404~(2006). 


\bibitem{ssnm2002}
S.~Savel'ev and F.~Nori, Nat. Mater. {\bf 1}, 179 (2002). 

\bibitem{ssfmfn}
S.~Savel'ev, F.~Marchesoni, and F.~Nori, 
Phys. Rev. Lett. {\bf 91}, 010601 (2003). 

\bibitem{ssfmfnPRE2004}
S.~Savel'ev, F.~Marchesoni, and F.~Nori, 
Phys. Rev. E {\bf 70}, 061107 (2004). 

\bibitem{ssfmfnPRE2005}
S.~Savel'ev, F.~Marchesoni, and F.~Nori, 
Phys. Rev. E {\bf 71}, 011107 (2005). 

\bibitem{ssfmfn2004}
S.~Savel'ev, F.~Marchesoni, and F.~Nori, 
Phys. Rev. Lett. {\bf 92}, 160602 (2004); 

\bibitem{wyangPRE79} 
W. Yang, K. Nelissen, M. Kong, Z. Zeng, and F. M. Peeters, Phys. Rev. E {\bf 79}, 041406 (2009). 

\bibitem{engel} 
A.~Engel, H.~W.~M\"{u}ller, P.~Reimann, and A.~Jung, 
Phys. Rev. Lett. {\bf 91}, 060602 (2003). 

\bibitem{ssPRB2004} 
S.~Savel'ev, V.~R.~Misko, F.~Marchesoni and F.~Nori, 
Phys. Rev. B~{\bf 71}, 214303~(2005). 

\bibitem{ertas} 
D.~Ertas, Phys. Rev. Lett. {\bf 80}, 1548 (1998). 

\bibitem{huangnb} 
L.~R.~Huang, J.~O.~Tegenfeldt, J.~J.~Kraeft, J.~C.~Sturm, 
R.~H.~Austin, and E.~C.~Cox, 
Nat. Biotechnol. {\bf 20}, 1048 (2002). 

\bibitem{berger} 
M.~Berger, J.~Castelino, R.~Huang, M.~Shah, and R.~H.~Austin,
Electrophoresis {\bf 22}, 3883 (2001). 

\bibitem{boxer} 
A.~Oudenaarden and S.~G.~Boxer, Science {\bf 285}, 1046 (1999). 

\bibitem{huangsc} 
L.~R.~Huang, E.~C.~Cox, R.~H.~Austin, and J.~C.~Sturm, 
Science {\bf 304}, 987 (2004). 

\bibitem{Gnann} 
M.~V.~Gnann, I.~Gazuz, A.~M.~Puertas, M.~Fuchs, and 
Th.~Voigtmann, Soft Matter {\bf 7}, 1980 (2011). 

\bibitem{Ebbens} 
S. J. Ebbens and J. R. Howse, Soft Matter {\bf 6}, 726 (2010). 

\bibitem{Buttinoni} 
I. Buttinoni, G. Volpe, F. K\"{u}mmel, and C. Bechinger, 
Submitted to: J. Phys.: Condens. Matter; e-print arXiv:1110.2202v3~(2012). 

\bibitem{ArXivSwimmer} 
G. Volpe, I. Buttinoni, D. Vogt, H.-J. K\"{u}mmerer, and C. Bechinger, Soft Matter {\bf 7}, 8810 (2011). 

\bibitem{chen} 
Q.~Chen, J.~K.~Whitmer, S.~Jiang, S.~C.~Bae, E.~Luijten, 
and S.~Granick, 
Science {\bf 331}, 199 (2011). 

\bibitem{SynytskaAFM}
S. Berger, L. Ionov, A. Synytska, Adv. Funct. Mater. {\bf 21}, 2338 (2011). 

\bibitem{SynytskaTex}
A. Synytska, R. Khanum, L. Ionov, C. Cherif, and C. Bellmann,
ACS Appl. Mater. Interfaces {\bf 3}, 1216 (2011). 

\bibitem{Kline} 
T. R. Kline, W. F. Paxton, T. E. Mallouk, and A. Sen, 
Angew. Chem. Int. Ed. {\bf 44}, 744 (2005). 

\bibitem{Tierno} 
P. Tierno, R. Albalat, and F. Sagu\'{e}s, 
Small {\bf 6}, 1749 (2010). 

\bibitem{KwintenEPL74} 
K. Nelissen, B. Partoens, I. Schweigert and F. M. Peeters, EuroPhys. Lett.~{\bf 74}, 1046~(2006).

\bibitem{BradyPRL} 
U. M. C\'{o}rdova-Figueroa and J. F. Brady,
Phys. Rev. Lett. {\bf 100}, 158303 (2008). 

\bibitem{BradyPRL2009} 
U. M. C\'{o}rdova-Figueroa and J. F. Brady,
Phys. Rev. Lett. {\bf 103}, 079802 (2009). 

\bibitem{bacteria} 
M. B. Wan, C. J. Olson Reichhardt, Z. Nussinov, and 
C. Reichhardt, 
Phys. Rev. Lett. {\bf 101}, 018102 (2008). 

\bibitem{Lowen1} 
B. ten Hagen, S. van Teeffelen, and H L\"{o}wen, 
J. Phys.: Condens. Matter {\bf 23}, 194119 (2011). 

\bibitem{Lowen2} 
S. van Teeffelen and H L\"{o}wen, 
Phys. Rev. E {\bf 78}, 020101(R) (2008). 

\bibitem{rotation} 
Taking into account rotational motion of a MS would lead to additional randomness (which is already present, due to the thermal stochastic term in Eq. (1)), in the trajectory of the MS. However, the preferable (average) direction of motion of the MS still can be controlled by employing illumination patterns. 

\bibitem{VRMCorbino} 
V. R. Misko and F. M. Peeters, 
Phys. Rev. B {\bf 74}, 174507 (2006). 

\bibitem{VRMCorbino1} 
N. S. Lin, V. R. Misko, and F. M. Peeters, 
Phys. Rev. Lett. {\bf 102}, 197003 (2009). 

\end{thebibliography}
\end{document}